\documentclass{amsart}
\usepackage{amsfonts}
\usepackage{graphicx}

\newtheorem{theorem}{Theorem}
\theoremstyle{plain}

\numberwithin{equation}{section}
\input{tcilatex}

\begin{document}
\title[Filtering Quantum White Noise]{Quantum Filtering of Markov Signals
with White Quantum Noise}
\author{V. P. Belavkin}
\address{Moscow Institute of Electronics and Mathematics\\
Moscow 109028 USSR}
\email{vpb@maths.nott.ac.uk}
\urladdr{http://www.maths.nott.ac.uk/personal/vpb/}
\thanks{}
\date{January 20, 1979}
\dedicatory{}
\thanks{Translated from Russian by R.V. Belavkin for Proceedings of an
International Workshop, held July 1994, in Nottingham, England: "Quantum
Communications and Measurement\textquotedblright , Plenum Press, New York
and London, pp. 381 -- 391. The original was published in: Radiotechnika i
Elektronika, 25 (1980) pp. 1445--1453.}

\begin{abstract}
Time-continuous non-anticipating quantum processes of nondemolition
measurements are introduced as the dynamical realizations of the causal
quasi-measurements, which are described in this paper by the adapted
operator-valued probability measures on the trajectory spaces of the
generalized temporal observations in quantum open systems. In particular,
the notion of physically realizable quantum filter is defined and the
problem of its optimization to obtain the best a posteriori quantum state is
considered. It is proved that the optimal filtering of a quantum Markovian
Gaussian signal with the Gaussian white quantum noise is described as a
coherent Markovian linear filter generalizing the classical Kalman filter.
As an example, the problem of optimal measurement of complex amplitude for a
quantum Markovian open oscillator, loaded to a quantum wave communication
line, is considered and solved.
\end{abstract}

\maketitle

\section{Introduction}

\typeout{Introduction}

At present, due to invention of laser as a coherent quantum generator, the
dynamical problem of optimal reception of optical signals with quantum
noise, taking into account the fundamental limitations caused by the quantum
nature of the electromagnetic waves, is becoming actual. Quantum information
and communication theory which has been developing so far (See recent review 
\cite{bib:1} and the references in there) is based on static (nontemporal),
or single step (instantaneous) theory of quantum measurement, and quantum
statistical inference does not take into account physical\ causality in due
coarse of the dynamical propagation. As the result, the optimal estimators
based on nontemporal quantum measurements are noncausal since they may
depend at each intermediate time $t$ not only on the past but also on future
results of measurements. An attempt to develop the temporal, multistep
variant of quantum measurement and decision theory to solve the problems of
the dynamical filtering in discrete time was made in \cite{bib:2}. However,
because of the very restrictive class of quantum measurement process as a
sequences of independent, nonadaptive quantum measurements considered in
this paper, even the simplest Gaussian problem for quantum linear filtering
(quantum Kalman filtering) was not solved, with the exception of a
degenerated case, and no time-continuous generalization is possible within
this restricted method.

Here we show, while solving the filtering problems for a quantum diffusion,
that these difficulties can be avoided if a natural much wider class of
temporal adaptive quantum measurements is considered, assuming their
dependence not only on time, but also on the results of preceding
measurements. Our dynamical model for the temporal quantum causal
measurements has a natural time-continuous setup based on quantum stochastic
differential equations. An appropriate quantum measurement device, that
realizes the adaptive measurements, is called the nonanticipating (or
causal) quantum filter with a memory. The optimization of estimation based
on independent quantum measurements in the narrower class of filters without
memory is in our treatment a problem with constraints leading to a poorer
quality of filtration. As is proved below, the optimal time-continuous
filtering of Markov Gaussian signals with the background quantum white noise
is realized by a quantum linear filter based on the adaptive coherent
quantum measurements. The optimal filter has a Markov memory such that the
next optimal quantum measurement in general depends on the result of the
present measurement but is independent of the preceding measurement results
in this Markovian case. We show that such coherent quantum filter can be
realized by indirect heterodyne measurements adding an independent vacuum
quantum noise, and by causal processing of the temporal measurement results
using a classical linear Kalman-Bucy filter \cite{bib:3}. A nonlinear (and
quasi-linear) generalization of this result for the non-Gaussian case, which
corresponds to the classical nonlinear filtering theory of Stratonovich \cite%
{bib:4}, is also of our interest an will be considered elsewhere.

Let us first consider a physical model of quantum open system with output
channel, in which the problem of optimal quantum filtering arises naturally,
giving a solution of this problem. Then we shall set up a general rigorous
formulation of the filtering problem as the problem of quantum optimal
temporal estimation, and derive the solution of this problem in the Gaussian
case.

\section{Optimal observation of quantum oscillator at the output of waveline}

\label{sec:1}

\typeout{Optimal observation of quantum oscilator at the output of waveline}

Let us consider an electromagnetic oscillator with frequency $\Omega $
loaded on a pair of communication channels, the transmission lines which are
described by wave conductivity $G$ and wave resistance $R$ respectively, as
shown in the figure, where $\Omega =1/\sqrt{LC}$, $G=\sqrt{c_{1}/l_{1}}$, $R=%
\sqrt{l_{2}/c_{2}}$, and propagation velocities of the waves $\upsilon $$%
_{i}=1/\sqrt{l_{i}c_{i}}$, $i=1,2$ are assumed to be equal: $\upsilon
_{1}=\upsilon _{2}=\upsilon $ :

\begin{center}
\includegraphics[height=5.5cm]{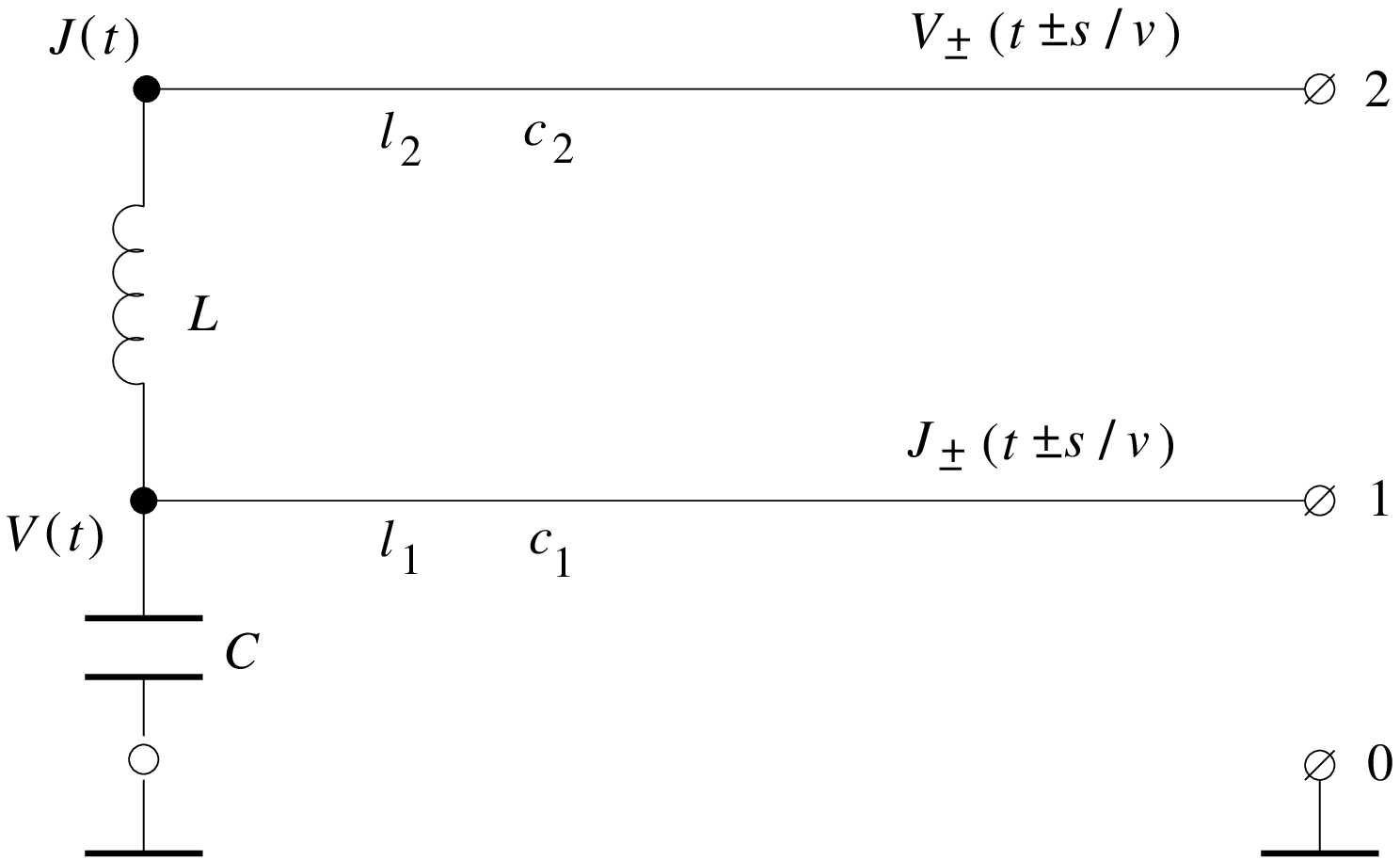}
\end{center}

We consider the case when the lines are homogeneous and that the measuring
device (receiver) which is set at their outputs of the lines $(0-1)$ and $%
(0-2)$ is ideally conjugated with these lines such that there is no
reflection of the received information (i.e. incoming waves $J_{-},V_{-}$)
into the radiated noise (i.e. outgoing waves $J_{+},V_{+}$) at the end of
the lines. Such system is open, but it is entirely described by the dynamic
variables of voltage $V(t)$ and current $J(t)$ on the contour $(L,C)$, and
by the pairs of running waves 
\begin{equation*}
J_{\pm }(t\pm s/\upsilon )=J_{1}(s,t)\pm GV_{1}(s,t)\,,\quad V_{\pm }(t\pm
s/\upsilon )=V_{2}(s,t)\pm RJ_{2}(s,t)
\end{equation*}%
of current and voltage in the first and second lines respectively as the
solutions to the telegraph wave equations. The boundary conditions $%
V_{1}(0,t)=V(t)$, $J_{2}(0,t)=J(t)$ induce on the open oscillator the
following pair of Langevin equations 
\begin{eqnarray*}
C\dot{V}(t)+GV(t)-J(t)=J_{+}(t)\,, &\quad &L\dot{J}(t)+RJ(t)+V(t)=V_{+}(t)\,,
\\
J_{-}(t)=J_{+}(t)-2GV(t)\,, &\quad &V_{-}(t)=V_{+}(t)-2RJ(t)\,,
\end{eqnarray*}%
where the second pair of equations, corresponding to these boundary
conditions, determines the output waves running from the oscillator to the
receiver.

Assuming, for simplicity, that the following gage invariance condition $%
R/L=G/C\equiv \gamma $ is fulfilled, the above system of equations can be
written in the one-dimension complexified form 
\begin{equation*}
\dot{x}(t)+\biggl(i\Omega +\frac{1}{2}\,\gamma \biggr)x(t)=x_{+}(t)\,,\quad
x_{-}(t)=x_{+}(t)-\gamma x(t)
\end{equation*}%
in terms of the complex amplitudes 
\begin{eqnarray*}
x(t) &=&\frac{1}{\sqrt{2}}\Big(\sqrt{C}V(t)+\mathrm{i}\sqrt{L}J(t)\Big)\,, \\
x_{\pm }(t) &=&\frac{1}{\sqrt{2}}\Big(\frac{1}{\sqrt{C}}J_{\pm }(t)+\frac{%
\mathrm{i}}{\sqrt{L}}V_{\pm }(t)\Big)\,.
\end{eqnarray*}

Now we move from the Langevin towards a stochastic description of the
corresponding quantum system. In the classical case, assuming that the input
noise $x_{+}\left( t\right) $, radiated by the receiver at the time $t-s/v$,
is the equilibrium noise of the temperature $T$, we can treat it as a
complex circular-invariant white noise with the intensity given by the
Nyquist formula $\sigma =kT$ (where $k$ is the Boltzman constant). In the
quantum case complex amplitude $x(t)$ should be replaced by the
corresponding operator $\check{x}(t)$ in the Heisenberg picture, satisfying
the canonical commutation relations 
\begin{equation}
\lbrack \check{x}(t),\check{x}(t)^{\ast }]\equiv \check{x}(t)\check{x}%
(t)^{\ast }-\check{x}(t)^{\ast }\check{x}(t)=\hbar \check{1}\,,\;\;
\label{eq:1}
\end{equation}%
where $\check{x}(t)^{\ast }$ is the Hermitian adjoint operator and $\check{1}
$ is the unit operator in the Hilbert space of quantum-mechanical states of
this system ($\hbar $ is the Planck constant, $\check{x}(t)/\sqrt{\hbar }=%
\check{a}(t)$ is the annihilation operator for oscillator quanta at the time 
$t$). By virtue of linearity the corresponding Heisenberg equation for $%
\check{x}(t)$ should have the same form as for the classical complex
amplitude $x(t)$, where the amplitude of the propagating wave $x_{+}$ have
to be replaced by the operator $\check{x}_{+}(t)$ with the commutation
relations ensuring preservation of the canonical commutation relations (\ref%
{eq:1}). This can be achieved if we take 
\begin{equation}
\lbrack \check{x}_{\pm }(t),\check{x}_{\pm }(t^{\prime })]=0\,,\quad \lbrack 
\check{x}_{\pm }(t),\check{x}_{\pm }(t^{\prime })^{\ast }]=\gamma \hbar
\delta (t-t^{\prime })\,,  \label{eq:2}
\end{equation}%
where we have taken into account that the commutators for the output wave $%
\check{x}_{-}(t)$ should coincide with the commutators of \ $\check{x}%
_{+}(t) $ due to its linear relation with $\check{x}_{+}(t)$, as it is also
assumed that $\check{x}_{+}(t)$ commutates both with $x(r)$, $\check{x}%
_{-}(r)$ and with $x(r)^{\ast }$, $\check{x}_{-}(r)^{\ast }$ for $r\leq t$.
As shown in \cite{bib:6}, the above equations and commutation relations can
be obtained by a canonical quantization of any open oscillator in \ the
\textquotedblleft rotating wave\textquotedblright\ representation and the
narrow-band approximation. Introducing the notations 
\begin{equation}
\check{v}(t)=\check{x}_{+}(t)=-\hat{v}(t)\,,\quad \check{y}(t)=\check{x}%
_{-}(t-s/\upsilon )=-\hat{v}(t)  \label{eq:2a}
\end{equation}%
for the forward and backward waves at the input and output of the
transmission line of length $s$ respectively, let us write these equations
in the following standard form 
\begin{equation}
{\frac{\mathrm{d}}{\mathrm{d}t}}\,\check{x}(t)+\alpha \check{x}(t)=\check{v}%
(t)\,,\quad \hat{y}(t+s/\upsilon )=\gamma \check{x}(t)+\hat{v}(t)\,,
\label{eq:3}
\end{equation}%
where $\alpha =i\Omega +\gamma /2$. The first equation, rewritten in terms
of quantum stochastic differentials as 
\begin{equation*}
\mathrm{d}\check{x}(t)+\alpha \check{x}(t)\mathrm{d}t=\mathrm{d}\check{v}%
_{t},
\end{equation*}%
where $\check{v}_{t}=\int_{0}^{t}\check{v}(t)\mathrm{d}t$, can be easily
integrated. The solution 
\begin{equation*}
\check{x}(t)=e^{-\alpha t}\check{x}(0)+\int_{0}^{t}e^{\alpha (r-t)}\mathrm{d}%
\check{v}_{r}
\end{equation*}%
does not depend on the output wave $\hat{y}(t^{\prime })$ with $t^{\prime
}<t+s/\upsilon $ and it commutes with $\hat{y}(t^{\prime })$ as well as with 
$\hat{y}(t^{\prime })^{\ast }$ for all such $t^{\prime }$. This
commutativity, reflecting the ideal conjugacy between the output measurement
device and the transmission lines, will play an important role for the
causal prediction of $\check{x}(t)$, called the nondemolition condition.
This condition, together with Markovianity of $\check{x}(t)$ corresponding
to the assumption that $\check{v}\left( t\right) $ is a quantum white noise,
which is fulfilled under the narrow-band approximation for the equilibrium
state of any measurement device sending the quantum thermal noise wave $%
\check{v}(t)$ into the line towards the oscillator, significantly simplifies
the optimization problem of the continuous measurement in the quantum system
under the consideration.

Let us now consider the problem of nondemolition observation of the
amplitude operator $\check{x}(t)$ of the quantum oscillator at the output of
the quantum noisy channel described as above. One would like to obtain an
optimal estimate $x(t)$ of this amplitude as a classical stochastic
complex-valued process, being observed in a real time on the output of the
transmission line. The mean square optimization problem defining at each $t$
such directly measurable nonanticipating estimate $x(t)$ based on the
results of the previous, in general indirect quantum nondemolition
measurements, is the minimization problem of the mean quadratic error 
\begin{equation}
\Big\langle\Big(\check{x}(t)-x(t)\Big)^{\ast }\Big(\check{x}(t)-x(t)\Big)%
\Big\rangle\,.  \label{eq:4}
\end{equation}

The continuous in time quantum nondemolition measurement is defined by any
measurable non-anticipating transformation of the received quantum process $%
\hat{y}\left( t\right) $ into the classical complex-valued one $x\left(
t\right) $, which should be in general randomized in order to include also
indirect nondemolition measurements of the noncommuting $\hat{y}(t)^{\ast }$
and $\hat{y}(t)$. Identifying the classical process $\left\{ x\left(
t\right) \right\} $ with a family of normal compatible operators in an
extended Hilbert space, one can describe it at each $t>0$ by a quantum
stochastic functional $x\left( t,\cdot \right) $ of the past output
operators $\hat{y}(t^{\prime })^{\ast },\hat{y}(t^{\prime }),t^{\prime }\leq
t$ and also of some additional independent quantum noises randomizing if
necessary this estimate, with the operator value $x(t)$ which commutes with
all operators $x(t^{\prime })^{\ast },x(t^{\prime })$ even if $t^{\prime
}\geq t$. Note that such estimator $x(t)$ will not necessary commute with
all past estimated operators $\check{x}(t^{\prime },\check{x}(t^{\prime
})^{\ast }$, although due to the nondemolition property it will commute with
them and all their future if $t^{\prime }\geq t-s/\upsilon $. Such quantum
directly observable process $x(t)$, which is the closest in the mean square
sense (\ref{eq:4}) to the non-observed quantum stochastic process $\check{x}%
(t)$, can be considered as a dynamical realization of the optimal temporal
quasi-measurement (or estimation) of the noncommuting $\check{x}(t^{\prime },%
\check{x}(t^{\prime })^{\ast }$, based on the indirect observation of the
given output process $\hat{y}\left( t\right) $.

In order to demonstrate such optimal temporal quasi-measurement let us
consider the problem of quantum filtering in the Gaussian case when the
initial quantum oscillator state is a circular-Gaussian with zero
mathematical expectation $\langle \check{x}(0)\rangle =0$ and average number
of quanta $\Sigma \geq 0$: 
\begin{equation}
\langle \check{x}(0)^{2}\rangle =0\,,\quad \langle \check{x}(0)^{\ast }%
\check{x}(0)\rangle =\hbar \Sigma \,,  \label{eq:4+}
\end{equation}%
and the quantum noise $\check{v}(t)$ is also circular-Gaussian white noise: $%
\langle \check{v}(t)\rangle =0$, 
\begin{equation}
\langle \check{v}(t)\check{v}(t^{\prime })\rangle =0\,,\quad \langle \check{v%
}(t)^{\ast }\check{v}(t^{\prime })\rangle =\hbar \nu \gamma \delta
(t-t^{\prime })\,,  \label{eq:4++}
\end{equation}%
where $\nu =(\exp \{\hbar \gamma /kT\}-1)^{-1}$ is the average number of
quanta in an equilibrium state of the receiver with the temperature $T$. As
will be shown in this paper, the optimal estimate $x(t)$ of the operator $%
\check{x}(t)$ minimizing the quadratic criterion (\ref{eq:4}) is given by a
nonorthogonal Hermitian-positive Gaussian operator-valued measure which
describes the coherent quasimeasurement of an integral nonanticipating
transformation of the output process $\hat{y}$. It can be realized by the
direct measurement of the classical complex random process $x(t)$ described
by the linear Langevin equation 
\begin{equation}
{\frac{\mathrm{d}}{\mathrm{d}t}}\,x(t)+\alpha x(t)=\kappa (t)\Big(%
y(t)-\gamma x(t)\Big)\,  \label{eq:5}
\end{equation}%
with $x\left( 0\right) =\langle \check{x}(0)\rangle =0$. The input process $%
y(t)$ in this equation is a classical complex (i.e. commutative normal)
process given by the additive randomizing transformation $y\left( t\right) =%
\hat{y}(t)+\hat{w}(t)$ of the received quantum process $\hat{y}(t)$, where $%
\hat{w}(t)$ is the complex amplitude of an additional independent quantum
noise with the opposite commutators 
\begin{equation*}
\lbrack \hat{w}(t),\hat{w}(t^{\prime })]=0\,,\quad \lbrack \hat{w}(t),\hat{w}%
(t^{\prime })^{\ast }]=-\hbar \gamma \delta (t-t^{\prime }),
\end{equation*}%
zero expectations $\langle \hat{w}(t)\rangle =0$ and the minimal vacuum
state correlations 
\begin{equation}
\langle \hat{w}(t)\hat{w}(t^{\prime })\rangle =0\,,\quad \langle \hat{w}%
(t)^{\ast }\hat{w}(t^{\prime })\rangle =\hbar \gamma \delta (t-t^{\prime
})\,.  \label{eq:7}
\end{equation}%
Since $y\left( t\right) $ is equivalent to the classical $\delta $%
-correlated complex process of the intensity $\sigma =\hbar \left( \nu
+1\right) \gamma $,$\,$ this Langevin equation can be identified with the
classical Kalman-Bucy filter which is usually written in the stochastic
differential form as 
\begin{equation*}
\mathrm{d}x(t)+\alpha x(t)\mathrm{d}t=\kappa (t)\Big(\mathrm{d}y_{t}-\gamma
x(t)\mathrm{d}t\Big)\,,
\end{equation*}%
in terms of the continuous diffusive input process $y_{t}=\int_{0}^{t}y(t)%
\mathrm{d}t$. In this equation $\kappa (t)=\left( \Sigma (t)-\nu \right)
/(1+\nu )$\thinspace , $\Sigma (t)$ is the solution of the Riccati equation 
\begin{equation}
{\frac{\mathrm{d}}{\mathrm{d}t}}\Sigma (t)=\frac{\gamma }{1+\nu }\left( \nu
-\Sigma (t)\right) \left( 1+\Sigma (t)\right) \,,  \label{eq:6}
\end{equation}%
with $\Sigma (0)=\Sigma _{0}$ and the effective classical white noise
intensity $\sigma =\hbar \gamma (1-\exp \{-\hbar \gamma /kT\})^{-1}$ Thus
the randomized nondemolition quasimeasurement of $\check{x}\left( t\right) $
is realized by the direct observation of $x\left( t\right) $ in the result $%
y\left( t\right) $ of indirect temporal observation of $\hat{y}\left(
t\right) $ by the heterodyning processing $\hat{y}\left( t\right) \mapsto 
\hat{y}(t)+\hat{w}(t)$ \cite{bib:7}, where $\hat{w}(t)$ plays the role of
the reference quantum wave. The mean quadratic error~(\ref{eq:4}) for
optimal filtration in this case has the minimal value $\hbar \Sigma (t)$,
which is not zero even under zero temperature $T=0$ corresponding to $\sigma
=\hbar \gamma $, in the sharp contrast with the classical case when $\sigma
=0$ at $T=0$. This corresponds to adding into the measurement channel a
vacuum quantum noise which makes possible the heterodyne indirect
measurement of $\hat{y}\left( t\right) $ statistically equivalent to a
classical filtering of the complex white noise of the minimal intensity $%
\hbar \gamma >0$. Note that this conclusion remains also valid for the
quasi-classical oscillator corresponding to $\Sigma _{0}\gg 1$, being
indirectly observed in a quantum wave line. However in the infinite
temperature case $T\rightarrow \infty $ when $\sigma \gg 1$, the quantum
consideration does not produce a substantial increase of the optimal
filtring mean quadratic error compared with the one given by the classical
Kalman-Bucy filter.

\section{Causal quantum measurement processes and filters}

\label{sec:2}

\typeout{Causal quantum measurement processes and filters}

Since the sequences $x(t_{1}),\ldots ,x(t_{2})$ of the discrete-time
measurements for arbitrary $\left\{ t_{1}<\ldots <t_{n}\right\} $ completely
determine a continuous function $x(t)$ under the temporal observation, the
quantum measurement process can be statistically described by the multitime
probability distributions $\mathbf{P}(\mathrm{d}x(t_{1}),\ldots ,\mathrm{d}%
x(t_{n}))$, $n=1,2,\ldots $ which define the probability measure $\mathbf{P}%
\left( \mathrm{d}\{x(t)\}\right) $ as a projective limit on the space of the
trajectories $\{x(t)\}$. The statistical structure of quantum theory
requires the probability measure to be a linear form with respect to the
density operator $\hat{\varrho}$ of a quantum state, represented as 
\begin{equation}
\mathbf{P}\left( \mathrm{d}\{x(t)\}\right) =\mathrm{Tr}\,\hat{\varrho}\Pi
\left( \mathrm{d}\{x(t)\}\right)  \label{eq:8}
\end{equation}%
Here $\mathrm{Tr}\,$ means the Hilbert space trace, $\Pi \left( \mathrm{d}%
\{x(t)\}\right) \geq 0$ is a Hermitian-positive measure on the space of the
observable trajectories $\{x(t)\}$ with values in an output operator algebra 
$\mathcal{B}$. We call such $\Pi $ the operator-valued probability measure
(OPM) since it must be normalized to the identity operator $\int \Pi \left( 
\mathrm{d}\{x(t)\}\right) =\hat{1}$ due to the normalization of $\mathbf{P}$%
. This measure defines, in particular, the multitime OPM $\Pi (\mathrm{d}%
x(t_{1})\ldots \mathrm{d}x(t_{n}))$ for any finite $n=1,2,\ldots $ which
induce the probabilities of the time-discrete temporal observations as the
linear functions of $\hat{\varrho}$ according to~(\ref{eq:8}).

Every decomposition of the identity $\hat{1}=\int \Pi (\mathrm{d}\{x(t)\})$
into the positive operators $\Pi (\mathrm{d}\{x(t)\})$ corresponds,
according to Naymark's theorem \cite{bib:8}, to a measurement of some
compatible set of operators, which are observed in an \textquotedblleft
extended quantum system\textquotedblright . But only those measures $\Pi $
correspond to the measurement processes physically realizable in real time,
which satisfy the causality condition $\Pi ^{t}(\mathrm{d}x^{t})\in \mathcal{%
B}^{t}$, where $\Pi ^{t}$ is the measure $\Pi $ on the space of
\textquotedblleft reduced\textquotedblright\ realizations $%
x^{t}=\{x(s)\}_{s\leq t}$, and $\{\mathcal{B}^{t}\},t\geq 0$ is a
nondecreasing family of operator algebras $\mathcal{B}^{s}\subseteq \mathcal{%
B}^{t}$, defining the nondemolition observables up to the time instants $t$.
Typically $\mathcal{B}^{t}$ is generated by the subfamily $\hat{b}^{t}=\{%
\hat{b}(s):s\leq t\}$ of a given operator family $\{\hat{b}(t)\},t\geq 0$,
describing an output quantum stochastic process $\hat{b}(t)$ as in the
example of first Section where $\hat{b}(t)=\hat{y}(t)$.

In order to have predictable behavior of a quantum object described by the
Heisenberg operators $\left\{ \check{x}\left( t\right) \right\} $ generating
an algebra $\mathcal{A}$ under the indirect temporal measurements in the
algebra $\mathcal{B}$, the subalgebras $\mathcal{B}^{t}$ must satisfy only
the commutativity condition $\mathcal{B}^{t}\subseteq \mathcal{A}%
_{t}^{\prime }$ for each $t$ with respect to the algebras $\mathcal{A}_{t}$
generated by the present and future Heisenberg operators $\check{x}_{t}=\{%
\check{x}(s):s\geq t\}$ of the object such that there exist conditional
expectation on $\mathcal{A}_{t}\vee \mathcal{B}^{t}$ with respect to $%
\mathcal{B}^{t}$ for each $t$. The measurement process should be also
selfpredictable, for which we shall assume the existence of the conditional
OPM $\Pi _{s}^{t}(x^{s},\mathrm{d}x_{s}^{t})$ on $x_{s}^{t}=\{x(t^{\prime
})\}_{t^{\prime }\in (s,t]}$, describing by the formula (\ref{eq:8}) the
non-anticipating processes of observation from any moment $s<t$ up to $t$
with known results $x^{s}=\{x(t)\}_{t\leq s}$ of the preceding measurements.
They should satisfy the compatibility condition 
\begin{equation}
\Pi _{t_{0}}^{t_{1}}(x^{t_{0}},\mathrm{d}x_{t_{0}}^{t_{1}})\Pi
_{t_{1}}^{t_{2}}(x^{t_{1}},\mathrm{d}x_{t_{1}}^{t_{2}})=\Pi
_{t_{0}}^{t_{2}}(x^{t_{0}},\mathrm{d}x_{t_{0}}^{t_{2}})  \label{eq:9}
\end{equation}%
for all $t_{0}<t_{1}<t_{2}$ such that the Hermitian-positive operators $\Pi
_{s}^{t}(x^{s},\mathrm{d}x_{s}^{t})^{\ast }=\Pi _{s}^{t}(x^{s},\mathrm{d}%
x_{s}^{t})$ commute $\Pi ^{s}\left( \mathrm{d}x^{s}\right) \in \mathcal{B}%
^{s}$, having the values in the relative commutants $\mathcal{B}_{s}^{t}$ of
the output subalgebras $\mathcal{B}^{s}$ with respect to $\mathcal{B}^{t}$.
Such a compatible family $\{\Pi _{s}^{t}\}_{s<t}$ of the conditional OPM $%
\Pi _{s}^{t}(x^{s},\mathrm{d}x_{s}^{t})\in \mathcal{B}_{s}^{t}$, normalized
to the identity operator $\hat{1}$, will be called \emph{non-anticipating
quantum filter}. The measures $\Pi _{s}^{t}$, which are independent of $%
x^{s} $, correspond to the filters without memory. The Markovian filters
described by the conditional measures $\Pi _{s}^{t}$, which depend only on
the last preceding value $x(s)$, are the simplest filters with the memory.
In addition, the transition operator-valued measure $\Pi _{s}^{t}(x(s),%
\mathrm{d}x(t))$ defines, according to (\ref{eq:9}), the multitime
conditional measures 
\begin{eqnarray*}
&&\Pi (x(t_{0}),\mathrm{d}x(t_{1})\cdots \mathrm{d}x(t_{n-1})\mathrm{d}%
x(t_{n})) \\
&=&\Pi _{t_{0}}^{t_{1}}(x(t_{0}),\mathrm{d}x(t_{1}))\ldots \Pi
_{t_{n-1}}^{t_{n}}(x(t_{n-1}),\mathrm{d}x(t_{n})),
\end{eqnarray*}%
satisfying the operator-valued Smolukhowsky equation 
\begin{equation}
\int_{X_{t_{1}}}\Pi _{t_{0}}^{t_{1}}(x(t_{0}),\mathrm{d}x(t_{1}))\Pi
_{t_{1}}^{t_{2}}(x(t_{1}),\mathrm{d}x(t_{2}))=\Pi _{t_{0}}^{t_{2}}(x(t_{0}),%
\mathrm{d}x(t_{2})).  \label{eq:10}
\end{equation}

In the case of a linear measurable space $X$, for example a complex $n$%
-dimensional space $\mathbb{C}^{n}$, it is convenient to describe the
Markovian filters by transitional operator-valued characteristic functions
(OCF) defined as operator-valued Fourier integrals%
\begin{equation}
\Phi _{s}^{r}(\mathbf{x}(s),\mathbf{u}(t))=\int e^{\mathrm{i}(\mathbf{u}%
(t)^{+}\mathbf{x}(t)+\mathbf{x}(t)^{+}\mathbf{u}(t))}\Pi _{s}^{r}(\mathbf{x}%
(s),\mathrm{d}\mathbf{x}(t)),  \label{eq:11}
\end{equation}%
where $\mathbf{u}$, $\mathbf{x}$ are $n$-dimensional complex columns, and $%
\mathbf{x}^{+}$, $\mathbf{u}^{+}$ are their conjugate rows. The
normalization condition for $\Pi _{s}^{t}$ gives $\Phi _{s}^{t}(\mathbf{x}%
(s),0)=\widehat{1}$, and the condition of Hermitian positivity $\Pi
_{s}^{t}\geq 0$ implies Hermitian positive-definiteness of the
operator-matrix $[\Phi (\mathbf{x}(s),\mathbf{u}_{k}-\mathbf{u}_{l})]\geq 0$%
, where $\{\mathbf{u}_{i},i=1,2,\ldots \}$ is any finite collection of
vectors $\mathbf{u}_{i}\in \mathbb{C}^{n}$. The operators $\Phi _{s}^{t}$
commute on the non overlapping time intervals, and the Markovian condition (%
\ref{eq:9}) can be written in the following differential operator form 
\begin{equation}
\Phi _{t_{1}}^{t_{2}}\left( -\mathrm{i}\frac{\partial }{\partial \mathbf{u}%
(t_{1})},\mathbf{u}(t_{2})\right) \Phi _{t_{0}}^{t_{1}}(\mathbf{x}(t_{0}),%
\mathbf{u}(t_{1}))|_{u(t_{1})=0}=\Phi _{t_{0}}^{t_{2}}(\mathbf{x}(t_{0}),%
\mathbf{u}(t_{2})).  \label{eq:12}
\end{equation}%
Under certain regularity condition it can be proved that the opposite is
also true: for every family of regular operator-valued functions $\{\Phi
_{s}^{t},s<t\}$, which are continuous at $u(t)=0$, satisfying the above
normalization, Hermitian positive-definiteness and Markovianity condition (%
\ref{eq:12}), the exists unique Markovian filter having the OCF $\Phi
_{s}^{t}$.

As an example of a Markovian filter we can take a linear coherent filter
which is defined by OCF function 
\begin{equation}
\Phi _{s}^{t}(\mathbf{x}(s),\mathbf{u}(t))=e^{\mathrm{i}\mathbf{u}(t)^{+}%
\mathbf{\hat{x}}(t)}e^{\mathrm{i}\mathbf{\hat{x}}(t)^{+}\mathbf{u}(t)},
\label{eq:13}
\end{equation}%
where $\mathbf{\hat{x}}(t)$ is vector-column composed of the operators $\hat{%
x}_{1}(t),\ldots ,\hat{x}_{n}(t)$ and $\mathbf{\hat{x}}(t)^{+}$ is
vector-row composed of the adjoint operators $\hat{x}_{1}(t)^{\ast },\ldots ,%
\hat{x}_{n}(t)^{\ast }$. The operators $\mathbf{\hat{x}}(t)$ entering into
this anti-normal ordered expression for $\Phi _{s}^{t}$ are assumed to
satisfy the following linear quantum stochastic differential equation 
\begin{equation}
\frac{\mathrm{d}}{\mathrm{d}t}\mathbf{\hat{x}}(t)+B(t)\mathbf{\hat{x}}%
(t)=K(t)\mathbf{\hat{b}}(t),\qquad \mathbf{\hat{x}}(s)=\mathbf{x}(s)
\label{eq:14}
\end{equation}%
with $\mathbf{\hat{b}}(t)$ defined as $m$ -dimensional column composed of
the annihilation operators $\hat{b}_{1}(t),\ldots ,\hat{b}_{m}(t)$ of the
quantum output process satisfying the following canonical commutation
relations 
\begin{equation*}
\left[ \hat{b}_{k}(t),\hat{b}_{l}(t^{\prime })\right] =0,\quad \left[ \hat{b}%
_{k},(t),\hat{b}_{l}^{\ast }(t^{\prime })\right] =\delta _{kl}\delta
(t-t^{\prime }).
\end{equation*}
Here $B(t)=\left[ \beta _{kl}(t)\right] $, $K(t)=\left[ \kappa _{kl}(t)%
\right] $ are complex matrices of the size $n\times n$ and $n\times m$
respectively, which may continuously depend on time. It is not hard to
verify by integrating the equation (\ref{eq:14}) that the characteristic
operator-function (\ref{eq:13}) satisfies all the above mentioned
properties, including (\ref{eq:12}). In fact it can be easily seen that,
since the operators $\mathbf{\hat{x}}(r)$ are linear transformations from
the annihilation ones $\mathbf{\hat{b}}(t)$, that ( \ref{eq:13}) is the
characteristic operator-function of the operator-valued Dirac $\delta $%
-measure 
\begin{equation*}
\Pi _{s}^{t}\left( \mathbf{x}(s),\mathrm{d}\mathbf{x}(t)\right) =\mathcal{N}%
_{-}\{\delta \left( \mathbf{\hat{x}}(t),\mathrm{d}\mathbf{x}(t)\right) \}\,,
\end{equation*}%
which corresponds to the anti-normal ordering \textquotedblleft $\mathcal{N}%
_{-}$\textquotedblright , when the creation operators $\mathbf{\hat{b}}%
^{\ast }(t)$ act to the left before $\mathbf{\hat{b}}(t)$. It is well known
that such measure is generated by the coherent projectors and the Lebesgue
measure on the complex space $\mathbb{C}^{n}$. For example, for the case of
a fixed $\mathbf{x}(0)=\mathbf{0}$ the measure $\Pi (t,\mathrm{d}\mathbf{x}%
)\equiv \Pi _{0}^{t}(\mathbf{0},\mathrm{d}\mathbf{x})$ has the form 
\begin{equation}
\Pi (t,\mathrm{d}\mathbf{x})=|t,\mathbf{x}\rangle \langle t,\mathbf{x}|%
\mathrm{d}\lambda (t,\mathbf{x})\,,  \label{eq:17}
\end{equation}%
where $|t,\mathbf{x}\rangle $ are the normalized right eigen vectors

\begin{equation}
\mathbf{\hat{x}}(t)|t,\mathbf{x}\rangle =\mathbf{x}|t,\mathbf{x}\rangle
,\quad \mathbf{x}\in \mathbb{C}^{n}  \label{eq:18}
\end{equation}%
for the operators $\mathbf{\hat{x}}(t)$ defined by the equation ( \ref{eq:14}%
) with the initial condition $\mathbf{\hat{x}}(0)=\mathbf{0}$. The element $%
\mathrm{d}\lambda $ of the volume in $\mathbb{C}^{n}$, normalizing the
expression (\ref{eq:17}), is given by 
\begin{equation*}
\mathrm{d}\lambda (t,\mathbf{x})=\frac{1}{\det \left[ \pi C(t)\right] }\Pi
_{i=1}^{n}\mathrm{d}\func{Re}x_{i}\mathrm{d}\func{Re}x_{i}\,,
\end{equation*}%
where $C(t)$ is the Hermitian matrix of commutators $\left[ \hat{x}%
_{k}\left( t\right) ,\hat{x}_{l}(t)^{\ast }\right] =(C(t))_{kl}$ satisfying
the equation%
\begin{equation*}
\frac{\mathrm{d}}{\mathrm{d}t}C\left( t\right) +B\left( t\right) C\left(
t\right) +C\left( t\right) B\left( t\right) ^{+}=K\left( t\right) K\left(
t\right) ^{+}
\end{equation*}
with $C(0)=0$ corresponding to $\mathbf{\hat{x}}\left( 0\right) =\mathbf{0}$.

A physical realization of the above linear coherent quantum Markovian filter
as optimal quasi-measurement of the operators $\{\mathbf{\hat{x}}(t)\}$ is
given by a precise direct measurement of the Markovian process $\mathbf{x}%
(t) $ described by the linear classical stochastic differential equation 
\begin{equation}
\frac{\mathrm{d}}{\mathrm{d}t}\mathbf{x}(t)+B(t)\mathbf{x}(t)=K(t)\mathbf{b}%
(t)\,,\quad \mathbf{b}(t)=\mathbf{\hat{b}}(t)+\mathbf{\hat{c}}(t)\,.
\label{eq:19}
\end{equation}%
Here $\mathbf{\hat{c}}(t)$ is independent of $\mathbf{\widehat{b}}(t)$
quantum \textquotedblleft vacuum\textquotedblright\ vector noise with the
opposite commutators 
\begin{equation*}
\left[ \hat{c}_{k}(t),\hat{c}_{l}(t^{\prime })\right] =0\,,\quad \left[ \hat{%
c}_{k}(t),\hat{c}_{l}(t^{\prime })^{\ast }\right] =-\delta _{kl}\delta
(t-t^{\prime }),
\end{equation*}%
which make the components of the sum $\mathbf{b}(t)$ commuting and also
commuting with the components of $\mathbf{b(t)}^{\ast }$, and $\langle \hat{c%
}(t)\rangle =0$, 
\begin{equation*}
\langle \hat{c}_{k}(t)\hat{c}_{l}(t^{\prime })\rangle =0\,,\quad \langle 
\hat{c}_{k}(t)^{\ast }\check{c}_{l}(t^{\prime })\rangle =\delta _{kl}\delta
(t-t^{\prime })\,,
\end{equation*}%
and so $\langle \hat{c}_{k}(t)\hat{c}_{l}(t^{\prime })^{\ast }\rangle =0$.
Therefore the momenta of the process $\mathbf{x}(t)$ coincide with the
anti-normal momenta for $\mathbf{\hat{x}}(t)$, in particular $\langle 
\mathbf{\hat{x}}(t)\rangle =\langle \mathbf{x}(t)\rangle $, 
\begin{equation*}
\langle \hat{x}_{k}(t)\hat{x}_{l}(t)\rangle =\langle x_{k}(t)x_{l}(t)\rangle
\,,\quad \langle \hat{x}_{k}(t)\hat{x}_{l}(t)^{\ast }\rangle =\langle
x_{k}(t)x_{l}(t)^{\ast }\rangle \,.
\end{equation*}

\section{Quantum Gaussian Diffusion and Linear Filtering}

\label{sec:3}

\typeout{Quantum Gaussian State Diffusion and Filtering}

Let $\check{\mathbf{x}}\left( t\right) $ be an $n$-dimensional quantum
diffusive process with zero initial expectation $\langle \check{\mathbf{x}}%
(0)\rangle =0$ and the initial correlations 
\begin{equation}
\langle \check{x}_{k}(0)\check{x}_{l}(0)\rangle =0\,,\qquad \langle \check{x}%
_{k}(0)^{\ast }\check{x}_{l}(0)\rangle =(R_{0})_{lk},  \label{eq:20}
\end{equation}%
where $R_{0}$ is a given positive-definite matrix. The process is defined by
the stochastic equation 
\begin{equation}
\frac{\mathrm{d}}{\mathrm{d}t}\check{\mathbf{x}}(t)+A(t)\check{\mathbf{x}}%
(t)=J(t)\check{\mathbf{a}}(t),  \label{eq:21}
\end{equation}%
where $A(t)$, $J(t)$ are matrices of size $n\times n$ and $n\times m$
respectively, and $\check{\mathbf{a}}(t)$ is white quantum noise satisfying
canonical commutation relations 
\begin{equation}
\left[ \check{a}_{k}(t),\check{a}_{l}(t^{\prime })\right] =0,\quad \left[ 
\check{a}_{k},(t),\check{a}_{l}^{\ast }(t^{\prime })\right] =\delta
_{kl}\delta (t-t^{\prime })  \label{eq:2'}
\end{equation}
with zero expectations $\langle \check{\mathbf{a}}(t)\rangle =0$ and nonzero
normal correlation matrix $Q(t)$: 
\begin{equation}
\langle \check{a}_{k}(t)\check{a}_{l}(t^{\prime })\rangle =0\,,\quad \langle 
\check{a}_{k}(t)^{\ast }\check{a}_{l}(t^{\prime })\rangle =(Q(t)_{lk}\delta
(t-t^{\prime }).  \label{eq:22}
\end{equation}%
We consider also an output system described at the output of a quantum
linear noisy channel 
\begin{equation}
\mathbf{\hat{b}}(t)=F(t)\check{\mathbf{x}}(t)+\mathbf{\hat{a}}(t)
\label{eq:23}
\end{equation}%
by $m\times n$ matrix $F(t)$ and by $m$-dimensional quantum white noise $%
\mathbf{\hat{a}}(t)$ with canonical commutators 
\begin{equation}
\left[ \hat{a}_{k}(t),\hat{a}_{l}(t^{\prime })\right] =0\,,\quad \left[ \hat{%
a}_{k}(t),\hat{a}_{l}(t^{\prime })^{\ast }\right] =\delta _{kl}\delta
(t-t^{\prime })\,,  \label{eq:24}
\end{equation}%
zero expectations $\langle \mathbf{\hat{a}}(t)\rangle =0$ and a nonzero
normal correlation matrix $N(t)$: 
\begin{equation}
\langle \hat{a}_{k}(t)\hat{a}_{l}(t^{\prime })\rangle =0\,,\quad \langle 
\hat{a}_{k}(t)^{\ast }\hat{a}_{l}(t^{\prime })\rangle =(N(t))_{lk}\delta
(t-t^{\prime }).  \label{eq:25}
\end{equation}%
The pair $(\check{\mathbf{a}}(t),\mathbf{\hat{a}}(t))$ is assumed to be
independent of $\check{\mathbf{x}}(0)$ at least in the wide sense, having
zero correlations with $\check{\mathbf{x}}(0)^{\ast }$, but it can have
nonzero matrix $T(t)$ of the mutual normal correlations in 
\begin{equation}
\langle \check{a}_{k}(t)\hat{a}_{l}(t^{\prime })\rangle =0\,,\quad \langle 
\check{a}_{k}(t)^{\ast }\hat{a}_{l}(t^{\prime })\rangle =(T(t))_{kl}\delta
(t-t^{\prime }).  \label{eq:26}
\end{equation}%
This is because we cannot assume them independent but satisfying the
commutation relations 
\begin{equation}
\left[ \check{a}_{k}(t),\hat{a}_{l}(t^{\prime })\right] =0\,,\quad \left[ 
\check{a}_{k}(t),\hat{a}_{l}(t^{\prime })^{\ast }\right] =(D(t))_{kl}\delta
(t-t^{\prime }),  \label{eq:27}
\end{equation}%
which are necessary for the nondemolition condition of mutual commutativity
of all components $\mathbf{\hat{b}}(t^{\prime })$ with $\check{\mathbf{x}}%
(t)^{\ast }$ for any $t\geq t^{\prime }$. If $\check{\mathbf{x}}\left(
t\right) $ satisfies the commutation relations 
\begin{equation}
\left[ \check{x}_{k}(t),\check{x}_{l}(t)\right] =0\,,\qquad \left[ \check{x}%
_{k}(t),\check{x}_{l}(t)^{\ast }\right] =(C\left( t\right) )_{kl},
\label{eq:28}
\end{equation}%
the matrix $D$ is defined by the nondemolition condition as a solution to
the equation $JD+CF^{+}=0$ at each $t$ (As it occurred in the example
considered in the Section \ref{sec:1}.) Usually the commutation relations $%
C\left( t\right) $ of the object are preserved, $C\left( t\right) =C_{0}$,
i.e. $AC_{0}+C_{0}A^{+}=JJ^{+}$. Note, however, that the nonzero commutation
relations in (\ref{eq:2'}) are not necessary for this preservation and for
the nondemolition property 
\begin{equation*}
\lbrack \check{x}_{k}(t),\hat{y}_{l}(t^{\prime })]=0\,,\quad \lbrack \check{x%
}_{k}(t),\hat{b}_{l}(t^{\prime })^{\ast }]=0\;\;\;\forall t\geq t^{\prime }
\end{equation*}%
if the process $\check{\mathbf{x}}(t)$ has the degenerate matrix $C_{0}$,
e.g. if it is classical: $C_{0}=0$, in which case also $D=0$ for the
nondegenerate $J$.

The problem of optimal nonstationary quantum filtering consist of finding a
nonaticipating physically realizable measurement process, denoted by $%
\mathbf{x}(t)$, at the output of the quantum channel, minimizing at each
time $t$ the mean of the quadratic error 
\begin{equation}
\left\vert \mathbf{\check{x}}(t)-\mathbf{x}(t)\right\vert
^{2}:=\sum_{i=1}^{n}(\check{x}_{i}(t)-x_{i}(t))^{\ast }(\check{x}%
_{i}(t)-x_{i}(t)).  \label{eq:29}
\end{equation}

\begin{theorem}
\label{th:1} Let the initial quantum vector $\mathbf{\check{x}}(0)$ be
Gaussian with zero mathematical expectation and the correlations (\ref{eq:20}%
), the pair $(\mathbf{\check{a}}(t),\widehat{\mathbf{a}}(t))$ be also
Gaussian with zero expectations and with correlations (\ref{eq:22}), (\ref%
{eq:25}), (\ref{eq:26}), and $\widehat{\mathbf{a}}(t)$ satisfy the canonical
commutation relations (\ref{eq:24}). Then the optimal by the criterion (\ref%
{eq:29}) quantum filter for the quantum Markov Gaussian process $\mathbf{%
\check{x}}(t)$, defined by the linear quantum stochastic equation (\ref%
{eq:21}), is the Markovian coherent filter (\ref{eq:13}), described by the
linear quantum stochastic equation (\ref{eq:14}), where $\mathbf{x}(0)=0$
and 
\begin{equation}
B=A+KF,\quad K=(PF^{+}+JT^{+})(N+I)^{-1}.  \label{eq:30}
\end{equation}%
Here $I$ is the identity $m\times m$-matrix, and $P=P(t)$ is the a
posteriori correlation matrix, which is given by the solution of the Riccati
equation 
\begin{equation}
\frac{\mathrm{d}}{\mathrm{d}t}%
P+AP+PA^{+}+(PF^{+}+JT^{+})(N+I)^{-1}(FP+TJ)=JQJ^{+}  \label{eq:31}
\end{equation}%
with the initial condition $P(0)=R_{0}$. The minimal expectation value of
the quadratic error (\ref{eq:29}) at the time $t$ is defined by the trace of 
$P\left( t\right) $. (The dependence of all the matrices in (\ref{eq:30}), (%
\ref{eq:31}) on $t$ is omitted for brevity.)
\end{theorem}

\noindent \textsc{Proof\/}. Given the fact that the criterion for the point
filtering (\ref{eq:29}) depends on the value $\mathbf{x}(t)$ of the
observation $\{\mathbf{x}(t)\}$ only at the last instant $t$, the average
value of the quadratic error is defined only by the single-time
operator-valued measure $\Pi (t,\mathrm{d}\mathbf{x}(t))$. Therefore one can
find the solution $\Pi ^{0}(t,\mathrm{d}\mathbf{x}(t))$ of the statistical
problem \cite{bib:10} of the optimal observation independently for each $t$.
If it proves to be a single-time operator-valued measure for some physically
realizable measurement process $\Pi ^{0}(\mathrm{d}\{\mathbf{x}(t)\})$ in
the sense of the Section \ref{sec:2}, then the problem of optimal filtering
will be solved.

Using the stochastic calculus for normal ordered quantum Gaussian variables,
which was developed in \cite{bib:10}, one can obtain the following
representation for the average of the quadratic error (\ref{eq:29}) as a
function of the operator-valued measure $\Pi (t,\mathrm{d}\mathbf{x})$: 
\begin{equation}
\langle \left\vert \mathbf{\check{x}}(t)-\mathbf{x}(t)\right\vert
^{2}\rangle =\sum_{i=1}^{n}(P(t))_{ii}+\int \mathrm{Tr\,}\hat{R}(t,\mathbf{x}%
)\Pi (t,\mathrm{d}\mathbf{x}).  \label{eq:32}
\end{equation}%
Here $P(t)$ is a matrix satisfying the Riccati equation (\ref{eq:31}), $\hat{%
R}(t,\mathbf{x})$ is the operator-valued function 
\begin{equation}
\hat{R}(t,\mathbf{x})=\sum_{i=1}^{n}(\hat{x}_{i}(t)-x_{i})^{\ast }\hat{%
\varrho}(t)(\hat{x}_{i}(t)-x_{i}),  \label{eq:33}
\end{equation}%
of the operators $\mathbf{\hat{x}}(t)$ satisfying the equation (\ref{eq:14})
defined by the matrices (\ref{eq:30}) with the initial condition $\mathbf{%
\hat{x}}(0)=0$, and $\hat{\varrho}(t)$ is a Gaussian density operator for
the output process $\mathbf{\hat{x}}(t)$. This $\hat{\varrho}(t)$ can be
represented in the form of a normally ordered expression 
\begin{equation*}
\hat{\varrho}(t)=\det (G(R-P)^{-1})\mathcal{N}\left[ \exp \left\{ -\sum_{k,l}%
\hat{x}_{k}^{\ast }(t)(R-P)_{kl}^{-1}\hat{x}_{l}(t)\right\} \right] ,
\end{equation*}%
where $G(t)$, $R(t)$ are the solutions of the linear equations 
\begin{equation*}
\frac{\mathrm{d}}{\mathrm{d}t}G+BG+GB^{+}=KK^{+},\quad \frac{\mathrm{d}}{%
\mathrm{d}t}R+AR+RA^{+}=JQJ^{+}
\end{equation*}%
with the initial conditions $G(0)=0$ and $R(0)=R_{0}$ respectively.

The second element of the right-hand side in (\ref{eq:32}) is not negative
because of the Hermitian positivity of the operator (\ref{eq:33}), and it is
equal to zero on the coherent measure (\ref{eq:17}) defined by equation (\ref%
{eq:18} ) for $\mathbf{x}_{0}=0\mathrm{.}$ Hence, it is clear that the
coherent measurement, described by this measure, is optimal for all $t$.
Therefore, the coherent observation described by the OCF $\Phi \left( t,%
\mathbf{u}\right) =\Phi _{0}^{t}\left( \mathbf{0},\mathbf{u}\right) $ in (%
\ref{eq:13}) with the initial condition $\mathbf{\hat{x}}(0)=0$ is the
optimal measurement process. The Theorem is proved.

Note that the a posteriori state for the quantum diffusion $\mathbf{\check{x}%
}(t)$ is described by the Gaussian stochastic density operator%
\begin{equation*}
\check{\varrho}(t)=\det (G(P+G)^{-1})\mathcal{N}\left[ \exp \left\{ -\sum (%
\check{x}_{k}(t)-x_{k}(t))^{\ast }(P+G)_{kl}^{-1}(\check{x}%
_{l}-x_{l}(t))\right\} \right]
\end{equation*}%
given by the solution $\mathbf{x}\left( t\right) $ of the \ classical
stochastic diffusion equation (\ref{eq:19}) with $\mathbf{x}(0)=0$.

As example, let us consider the $n$-dimensional quantum oscillator (\ref%
{eq:21}) that has $\mathbf{\check{a}}(t)=\mathbf{\hat{a}}(t)$, $JJ^{+}=\hbar
(A+A^{+})$, such that the commutators 
\begin{equation}
\left[ \check{x}_{k},\check{x}_{l}\right] =0,\qquad \left[ \check{x}_{k},%
\check{x}_{l}^{\ast }\right] =\hbar \delta _{kl}  \label{eq:34}
\end{equation}%
are preserved. The matching $m$-dimensional quantum communication line is
described by the output wave 
\begin{equation}
\mathbf{\hat{y}}(t)=G(t)\mathbf{\check{x}}(t)-J(t)\mathbf{\hat{a}}(t).
\label{eq:35}
\end{equation}%
This wave commutes for $G=A+A^{+}$ with $\mathbf{\check{x}}(t^{\prime
})^{\ast }$, $t^{\prime }\geq t$. In the former dimensionless units this
line is described by the wave (\ref{eq:23}), where $F=-J^{+}/\hbar $, so
that $\mathbf{\hat{y}}(t)=\hbar F^{+}\mathbf{\hat{b}}(t)$. Let the
correlation matrix $N$ of the quantum channel noise $\mathbf{\hat{a}}(t)$ be
proportional to the identity matrix: $N=\nu I$. Then the equation (\ref%
{eq:31}) for the dimensionless a posteriori correlation matrix $S=P/\hbar $
will have the form 
\begin{equation}
\dot{S}=\mathrm{i}[S,H]-\frac{1}{2(\nu +1)}\left( (S+I)G(S-\nu I)+(S-\nu
I)G(S+I)\right) ,  \label{eq:36}
\end{equation}%
where $H=(A-A^{+})/2i$, and $K(t)=(\nu I-S)J/(1+\nu )$. In particular, in
the one-dimensional case $m=n=1$, denoting $S=\Sigma $, $G=\gamma $, we
obtain (\ref{eq:6}). It is interesting to note that the stationary solution $%
S=\nu I$ of the equation (\ref{eq:36}) corresponds to a nonstochastic
optimal filtering which does not require at all any observation since in
this case the amplification coefficient $K(t)$ is zero. The initial
condition $S_{0}=\nu I$ for the stationary solution $S\left( t\right) =S_{0} 
$ of the Riccati equation (\ref{eq:36}) in the case $H=\gamma I$ can be
interpreted as a condition of thermal equilibrium for the initial
statistical state of the oscillator having $H=\Omega $ in the one
dimensional case, and the the quantum channel noise equilibrium state as it
corresponds to equality of the temperature for the equilibrium state in the
communication line and of the temperature $T=\hbar \Omega /k\ln \left(
1+\Sigma ^{-1}\right) $ for the oscillator initial Gaussian with $\Sigma
_{0}=\gamma $.

\end{document}